\documentclass[11pt]{article}
\usepackage{moriond,epsfig}
\usepackage[english]{babel}
\usepackage{graphicx}
\usepackage{marvosym}

\def\be{\begin{equation}}
  \def\ee{\end{equation}}
\def\bea{\begin{eqnarray}}
  \def\eea{\end{eqnarray}}

\newcommand{\warn}[1]{}
\newcommand{\pt}{$p_{\mathrm{T}}$}

\begin{document}
\vspace*{0.3cm}
\title{Low Mass Dimuon Production in Indium-Indium Collisions at the CERN SPS}

\author{M.~Floris$^{1}$,
  R.~Arnaldi$^{9}$,
  R.~Averbeck$^{11}$,
  K.~Banicz$^{2,4}$,
  J.~Castor$^{3}$,
  B.~Chaurand$^{7}$,
  C.~Cicalo$^{1}$,
  A.~Colla$^{9}$,
  P.~Cortese$^{9}$,
  S.~Damjanovic$^{4}$,
  A.~David$^{2,5}$,
  A.~de~Falco$^{1}$,
  A.~Devaux$^{3}$,
  A.~Dress$^{11}$,
  L.~Ducroux$^{6}$,
  J.~Farjeix$^{3}$
  H.~En'yo$^{8}$,
  A.~Ferretti$^{9}$,
  A.~F\"orster$^{2}$,
  P.~Force$^{3}$,
  N.~Guettet$^{2,3}$,
  A.~Guichard$^{6}$,
  H.~Gulkanian$^{10}$,
  J.~Heuser$^{8}$,
  M.~Keil$^{2,4}$,
  L.~Kluberg$^{2,7}$,
  J.~Lozano$^{5}$,
  C.~Louren\c{c}o$^{2}$,
  F.~Manso$^{3}$,
  A.~Masoni$^{1}$,
  P.~Martins$^{2,5}$,
  A.~Neves$^{5}$,
  H.~Ohnishi$^{8}$,
  C.~Oppedisano$^{9}$,
  P.~Parracho$^{2}$,
  P.~Pillot$^{6}$,
  G.~Puddu$^{1}$,
  E.~Radermacher$^{2}$,
  P.~Ramalhete$^{2,5}$,
  P.~Rosinsky$^{2}$,
  E.~Scomparin$^{9}$,
  J.~Seixas$^{2,5}$,
  S.~Serci$^{1}$,
  R.~Shahoyan$^{2,5}$,
  P.~Sonderegger$^{5}$,
  H.J.~Specht$^{4}$,
  R.~Tieulent$^{6}$,
  G.~Usai$^{1}$,
  R.~Veenhof$^{2,5}$ and
  H.K.~W\"ohri$^{1,2}$\\
  (NA60 Collaboration)}

~

\address{
  $^{~1}$Univ.\ di Cagliari and INFN, Cagliari, Italy;
  $^{~2}$CERN, Geneva, Switzerland;
  $^{~3}$LPC, Univ.\ Blaise Pascal and CNRS-IN2P3, Clermont-Ferrand, France;
  $^{~4}$Univ.\ Heidelberg, Heidelberg, Germany;
  $^{~5}$IST-CFTP, Lisbon, Portugal;
  $^{~6}$IPN-Lyon, Univ.\ Claude Bernard Lyon-I and CNRS-IN2P3, Lyon, France;
  $^{~7}$LLR, Ecole Polytechnique and CNRS-IN2P3, Palaiseau, France;
  $^{~8}$RIKEN, Wako, Saitama, Japan;
  $^{~9}$Univ.\ di Torino and INFN, Italy;
  $^{10}$YerPhI, Yerevan, Armenia;\\
  $^{11}$SUNY, Stony Brook, NY, USA}

\maketitle\abstracts{NA60 is a fixed-target experiment at the CERN SPS
  which measured dimuon production in nucleus-nucleus and
  proton-nucleus collisions.  In this paper we report on a precision
  measurement of low-mass muon pairs in 158 AGeV indium-indium
  collisions. A significant excess of pairs is observed above the
  yield expected from neutral meson decays.  The excess can be
  isolated by subtraction of expected sources, thanks to the excellent
  mass resolution and large sample size.}

The ultimate goal of heavy ion collisions is the detection of
signatures of a phase transition from hadronic matter to a deconfined
and/or chirally restored medium.  One of the main
observation from the heavy ion program at the CERN SPS is the excess
of low mass dimuons in nucleus-nucleus collisions reported by the
CERES experiment~\cite{Agakichiev:mv,Agakichiev:1997au}.  This is
usually understood in terms of in-medium modification of the $\rho$
meson, which could convey information about chiral symmetry
restoration: changes in mass and/or width of the $\rho$ are predicted
as the chiral phase transition is approached. Most of the models agree
in predicting an increase in the width of the
$\rho$~\cite{Rapp:1999ej,Brown:2001nh} while several different
predictions were formulated for its mass.  The CERES data could be
described by several significantly different models, so that an
unambiguous interpretation of this result was not possible.  A clear
discrimination between different theoretical models requires good
statistics and mass resolution, together with a good signal to
background ratio.  These requirements are met by the NA60 experiment.

NA60 is a fixed target experiment which measures dimuon production in
nucleus-nucleus and proton-nucleus collisions at the CERN SPS. The
apparatus is composed of 4 main detectors:\emph{(i)} a muon
spectrometer, which tracks muons and provides the trigger signal to
the experiment; \emph{(ii)} a Zero Degree Calorimeter, for centrality
measurements; \emph{(iii)} a silicon vertex tracker, which tracks
particles in the vertex region; \emph{(iv)} a silicon beam tracker,
which tracks incoming beam particles before they hit the target. A
complete description of the apparatus can be found
elsewhere~\cite{Gluca:2005}. Here we only briefly describe the
detector concept. Experiments measuring dimuons usually put a hadron
absorber in front of the muon spectrometer, to make sure only muons
can reach this detector.  This is also the main limiting factor,
because of energy loss fluctuations and multiple scattering.  NA60
overcomes this limitations by means of a vertex
tracker~\cite{Keil:2005zq}, placed before the hadron absorber. Muons
reconstructed in the muon spectrometer are extrapolated back to the
vertex region and matched to vertex tracks, using both angles and
momenta. This leads to a much improved vertexing (the origin of the
muons can be accurately determined) and mass resolution (from roughly
80~MeV to 20~MeV at the $\phi$ mass).

\begin{figure}[tbp]
  \centering
  \includegraphics[width=5.5cm,height=4.5cm]{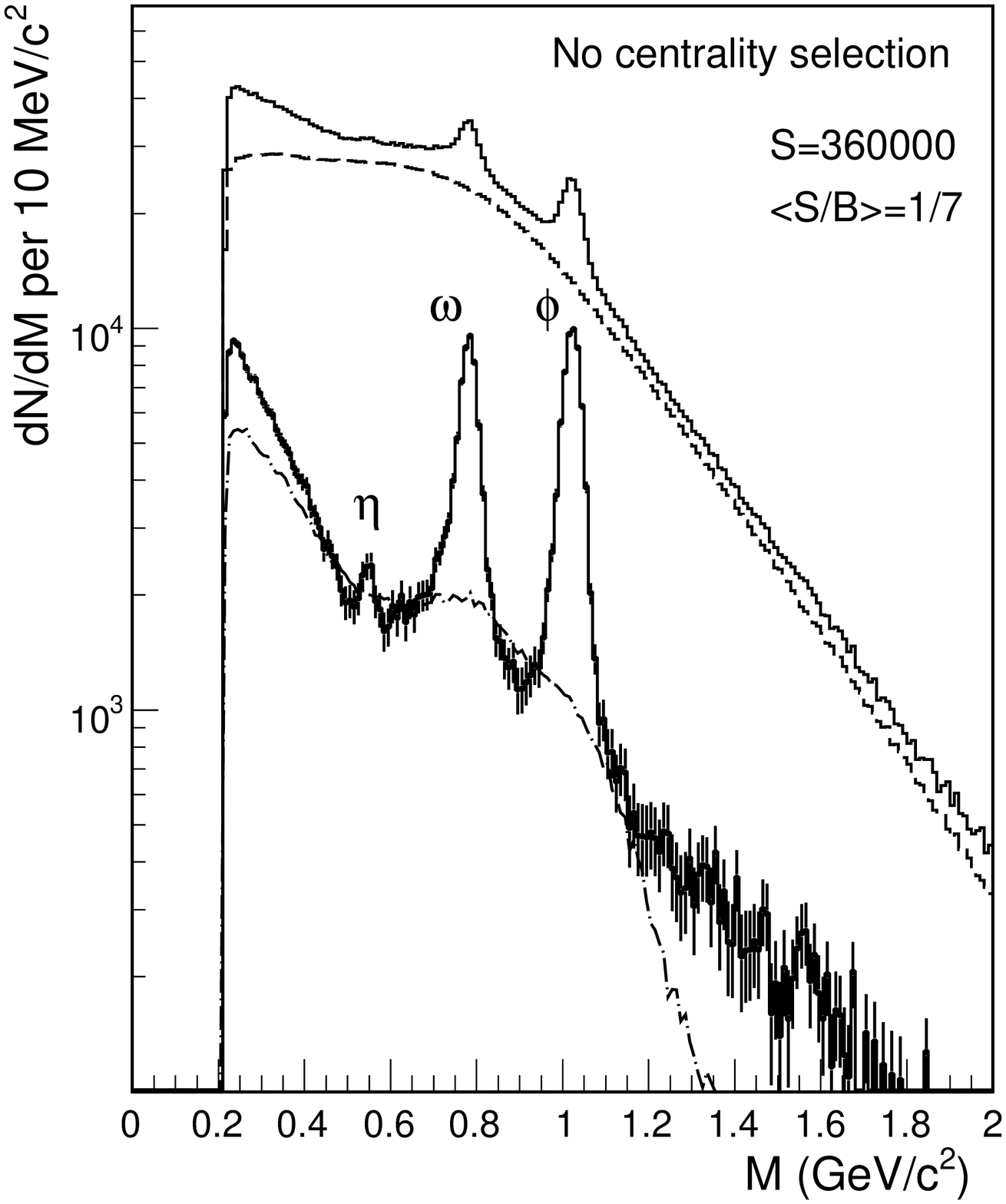}
  \hspace{1cm}
  \includegraphics[width=5.5cm,height=4.5cm]{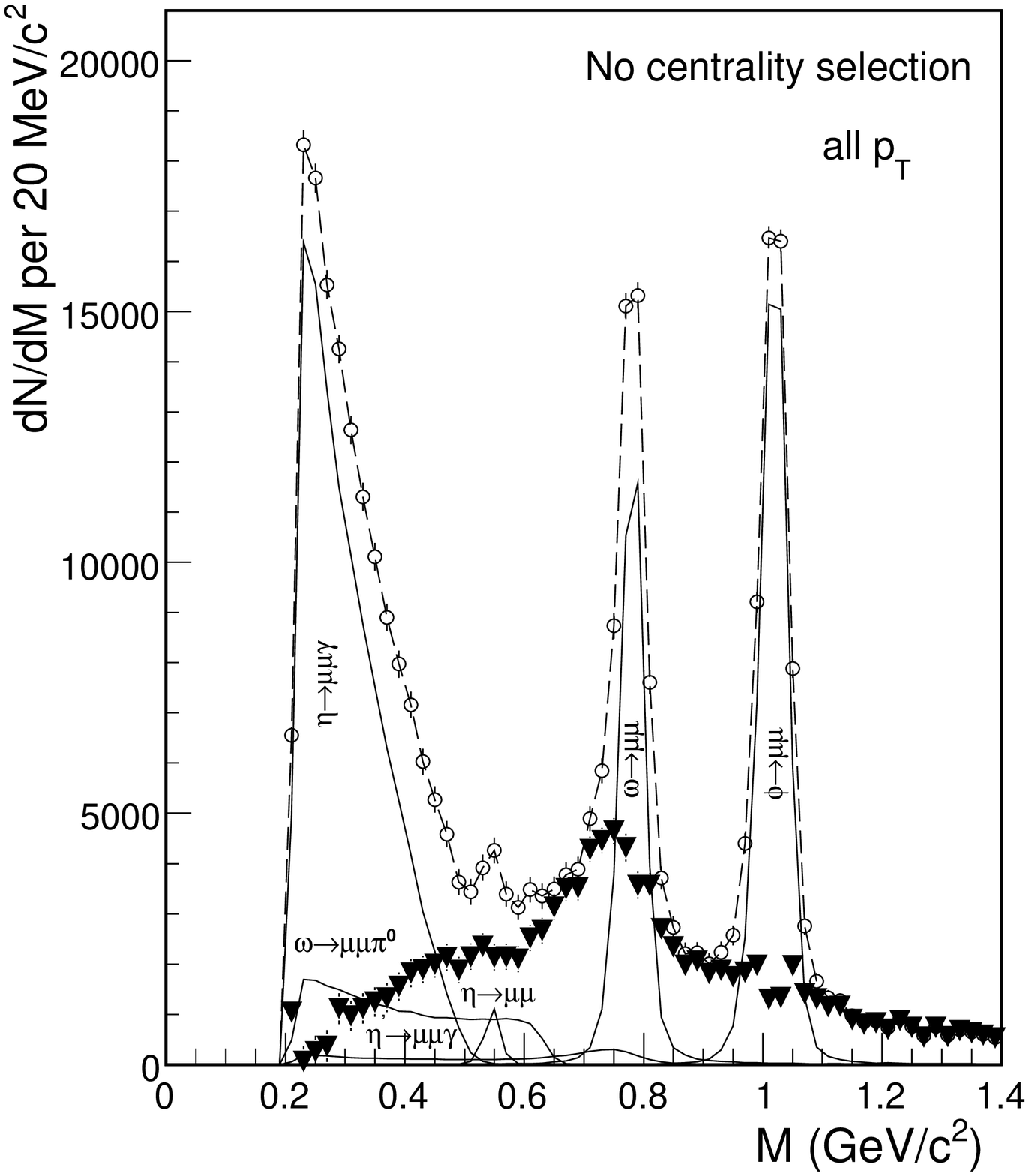}

  \caption{Mass spectrum after combinatorial background (dashed) and
    fake matches (dot-dashed) subtraction (left). Isolation of excess
    by subtraction of expected sources: total data (open symbols),
    expected sources (contious line), difference data (thick
    triangles), sum of cocktail sources and difference data (dashed).}
\label{fig:mass_spectra}
\end{figure}

NA60 physics programme includes topics from the three mass region in
which the dimuon invariant mass spectrum is customarily
divided~\cite{Lourenco:2001wi}. In this paper results on the Low Mass
Region (LMR, $\mathrm{M}<1.2~\mathrm{GeV}$) are reported. The data
were collected in indium-indium collisions in 2003; the sample
analyzed in this work consists of roughly half of the 230 million
events put on tape. The data were selected requiring only one vertex
in the target region, to avoid re-interactions and pile-up events.
The tracks reconstructed in the muon spectrometer were then matched to
vertex tracks as stated above, allowing to extract the invariant mass
spectrum of the resulting muon pairs. This is affected by two main
sources of background: the combinatorial background and the fake
matches, which have to be subtracted.  The combinatorial background is
the contribution to the dimuon spectrum of uncorrelated pairs coming
from decays of $\pi$ and K. This is estimated with an event mixing
technique: two muons from different events are randomly paired to
build an invariant mass spectrum, which is uncorrelated by
construction. This technique automatically takes into account all
experimental details, provided the muons are taken with the correct
normalization. The NA60 apparatus not only triggers opposite sign
pairs ($\mu^{+}\mu^{-}$) but also like sign pairs ($\mu^{-}\mu^{-}$
and $\mu^{+}\mu^{+}$), which are made of uncorrelated muons only. The
real and mixed spectra for the like-sign should then be identical.
This can be used to asses the quality of the estimated background. The
accuracy was found to be of the order of 1\% over several orders of
magnitude~\cite{Shahoian:2005zx}.  As for fake matches, they are
mistakes produced by the matching algorithm: a muon track can be
matched to the wrong track in the vertex telescope, giving rise to a
distorted spectrum. These can be estimated in two different ways.  The
first approach is an overlay Monte Carlo technique, where a Monte
Carlo dimuon is reconstructed on top of a real event, allowing to
check the probability of getting a fake match. The second method is an
event mixing method, which extracts the probability for fake matches
from data alone. The basic idea is to match the tracks in the muon
spectrometer from one event to the vertex tracks of a different event.
All the matches obtained in this way are fake by construction.  This
technique is more complicated but more rigorous. It is not used in the
present analysis, though is crucial for other NA60 physics
topics~\cite{Shahoian:2005zx}.  The two methods agree within 5\%. Fake
matches yield is less than 10\% of the combinatorial background. The
quality of the data after combinatorial background and fake matches
subtraction is very good (Fig.~\ref{fig:mass_spectra}, left): the
signal/background ratio is 1/7 and the mass resolution at the $\phi$
peak is 23 MeV. The analysis of the LMR data was done in 4 centrality
bins, selected using charged particle multiplicity density: peripheral
(4-30), semiperipheral (30- 110), semicentral (110-170) and central
(170-240).  The signal to background ratio, from peripheral to central
is: 2, 1/3, 1/8, 1/11.  The acceptance of NA60 extends all the way
down low masses and \pt, as opposed to previous dimuon experiments.
This is important because the excess observed by CERES in the
dielectron channel is stronger at low pt.

The first step in the analysis was the study of the peripheral bin. In
order to extract particle ratios in full phase space, data were fitted
with expected sources: $\eta \rightarrow \mu\mu$, $\omega \rightarrow
\mu\mu$, $\phi \rightarrow \mu\mu$, $\rho \rightarrow \mu\mu$, $\omega
\rightarrow \mu\mu\pi^0$, $\eta \rightarrow \mu\mu\gamma$,
$\eta^\prime \rightarrow \mu\mu\gamma$. The fit was done in 3
\pt~bins. The quality of the fit is very good.  The results are found
to be nearly independent of \pt~and in agreement with expectations,
confirming that the acceptance is well under control.  The particle
ratios obtained from the fits without \pt~selection, corrected for
acceptance and extrapolated to full phase space are
$\eta/\omega=0.88\pm0.03$, $\phi/\omega=0.094\pm0.004$ and
$\rho/\omega=1.62\pm0.10$.  The errors quoted are purely statistical.
The value for $\eta/\omega$ agrees with previously published
results~\cite{Agakichiev:mv}. The value for $\phi/\omega$ is higher
than measured in pp and pA collisions, as expected from strangeness
enhancement.  The value for $\rho/\omega$ is higher than expected from
pp or pA, probably due to the fact that even in peripheral
indium-indium collisions some hadronic medium is created, where the
$\rho$ is produced by $\pi\pi$ annihilation.
\begin{figure}[tbp]
  \centering
  \includegraphics[width=5cm]{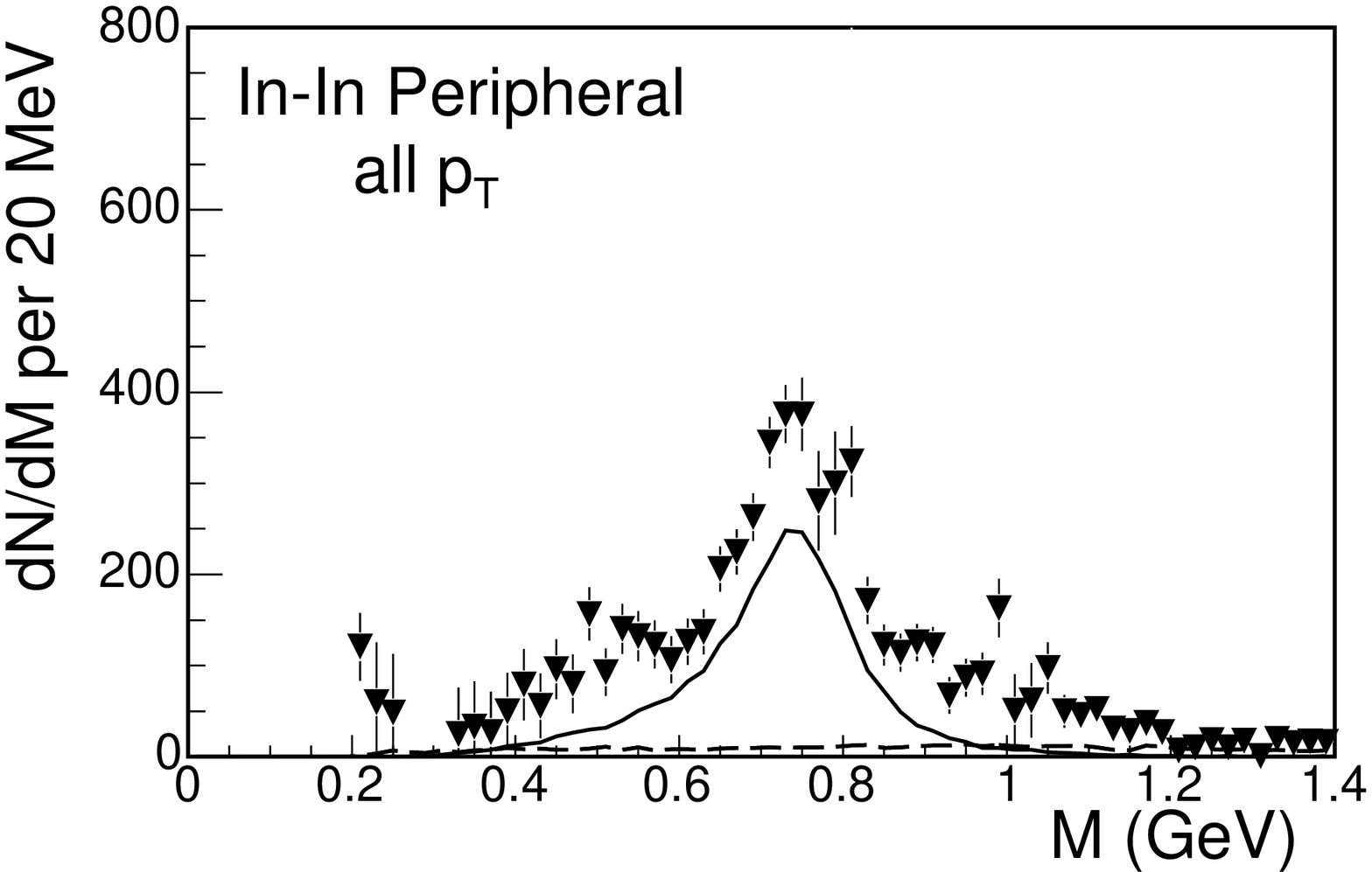}
  \hspace{1cm}
  \includegraphics[width=5cm]{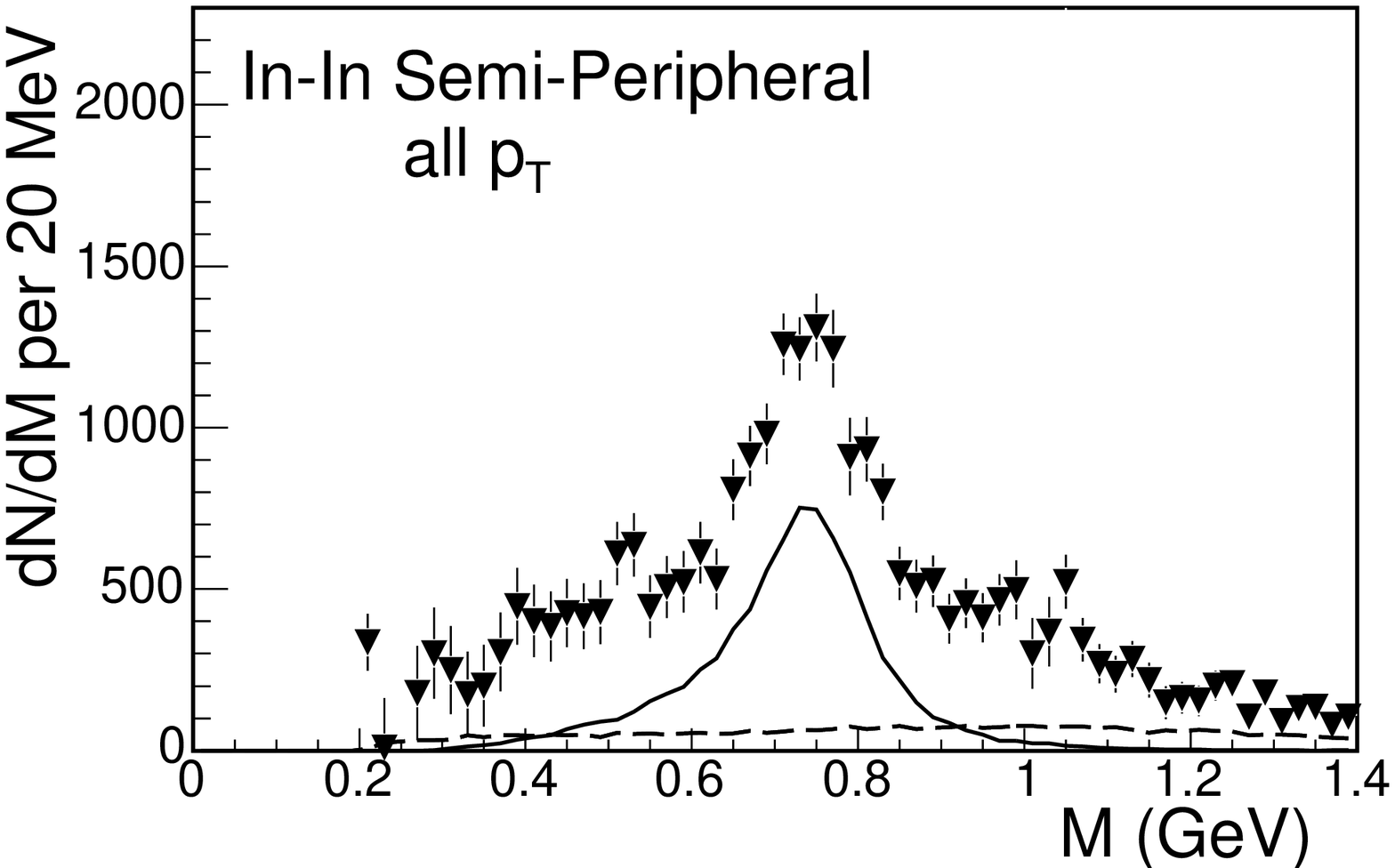}
  \\  
  \includegraphics[width=5cm]{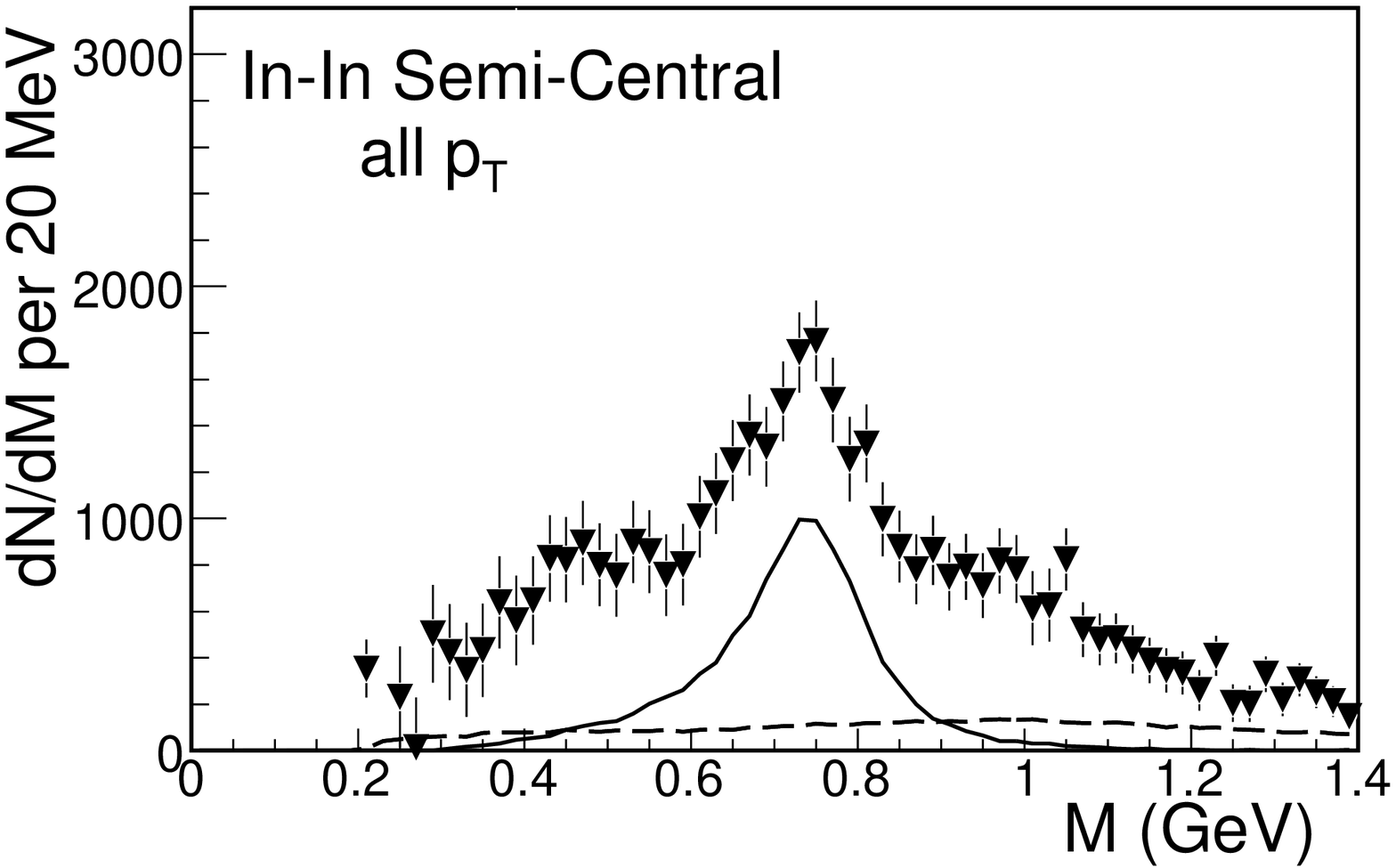}
  \hspace{1cm}
  \includegraphics[width=5cm]{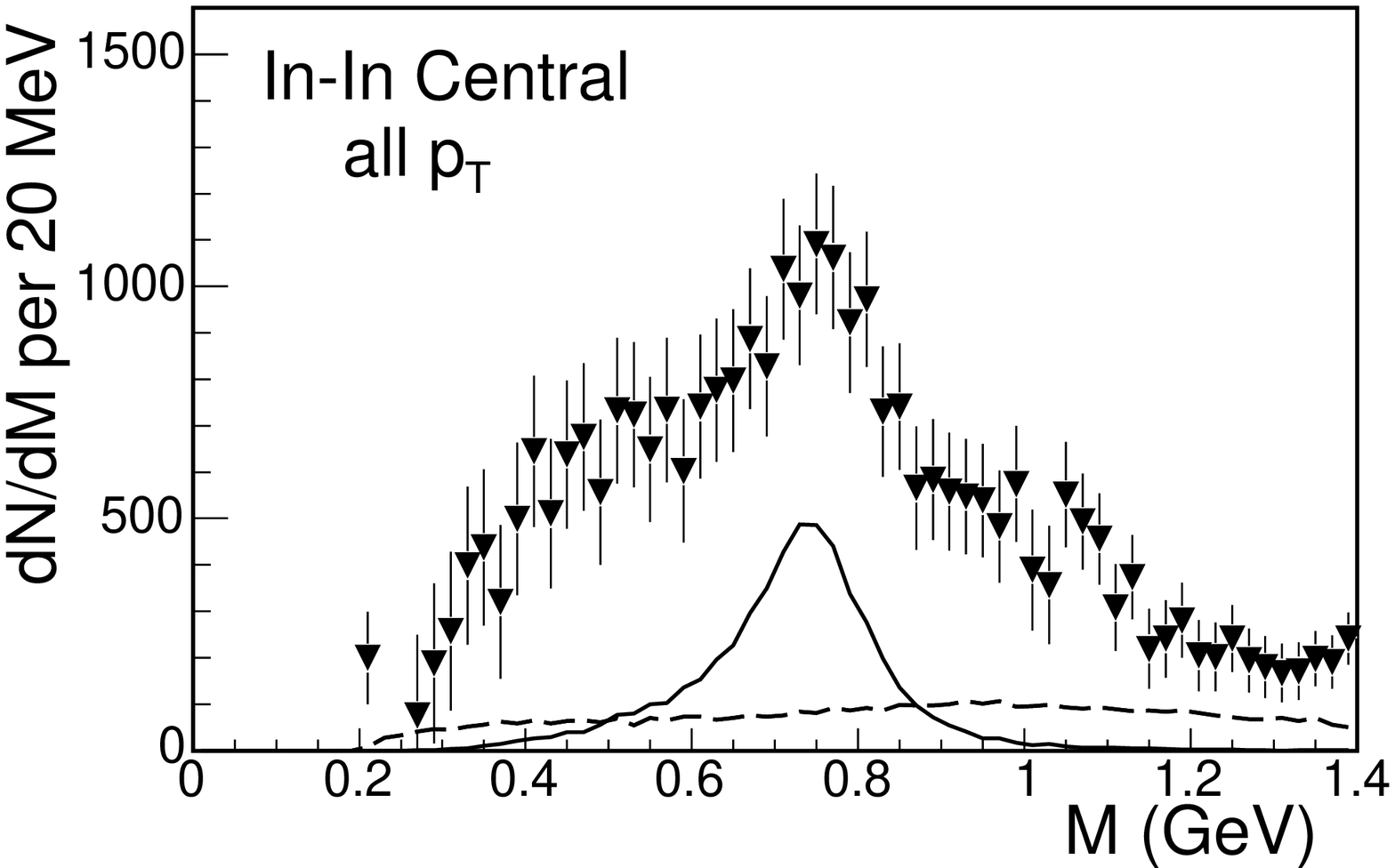}  
  \caption{Excess mass spectra. The cocktail (solid) and
    the level of uncorrelated charm decays (dashed) are shown for
    comparison. The errors are purely statistical.}
  \label{fig:excess_centrality}
\end{figure}

\begin{figure}[tbp]
  \centering
  \includegraphics[width=5cm,height=4.25cm]{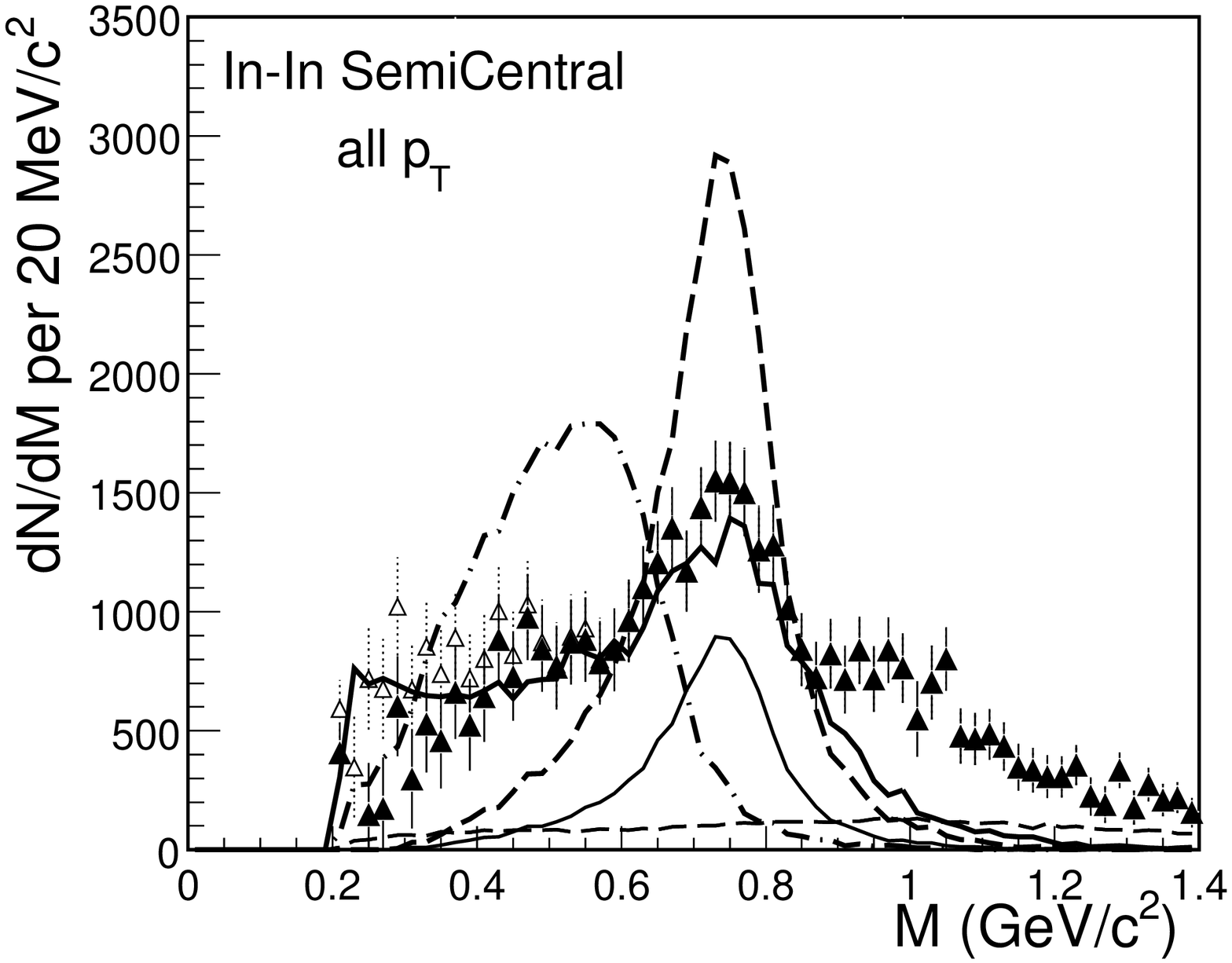}
  \hspace{1cm}
  \includegraphics[width=5cm,height=4.5cm]{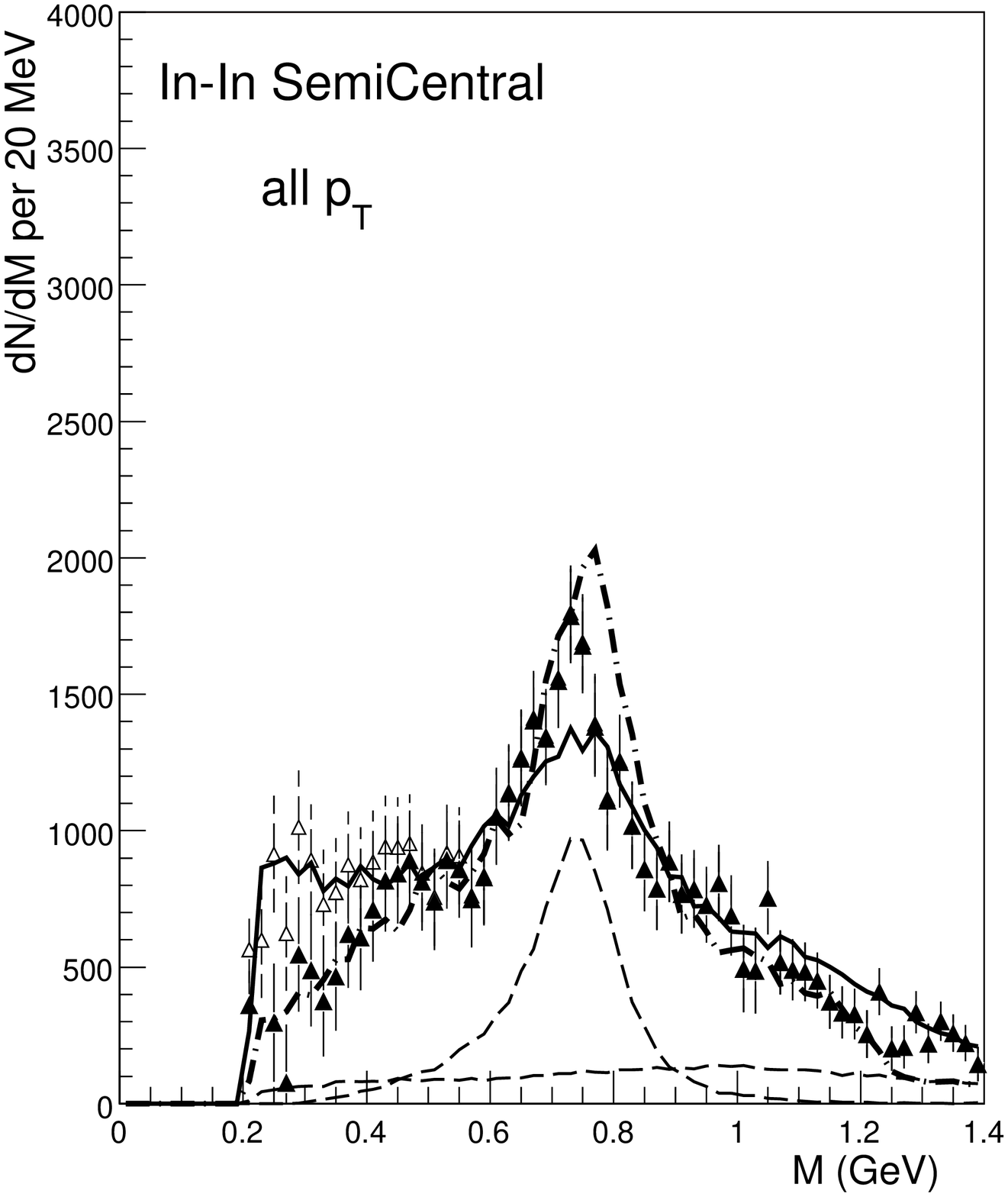}
  \caption{Comparison of the semi-peripheral excess mass spectrum to
    model predictions. The open data points show the difference
    spectrum obtained decreasing the $\eta$ yield by 10\%, reflecting the
    systematic uncertainty in this normalization. Left: Cocktail $\rho$
    (thin solid), unmodified $\rho$ (dashed), in-medium broadening~$^{3}$
    (thick solid), dropping mass~$^{4}$ (dashed-dotted). Right:
    Comparison to models of Ref.~$^{10}$ (solid) and Ref.~$^{9}$
    (dashed-dotted).}
  \label{fig:theory}
\end{figure}

The more central bins could not be studied by a simple fitting
procedure, because of the unknown source reported by CERES. Data were
compared to a conservative cocktail, defined as follows: \emph{(i)}
the yields of $\phi$ and $\omega$ is fixed from the 2-body-decay peaks
under the assumption that the underlying continuum is smooth (i.e. so
as not to create any bump or dip in the spectrum after subtraction of
these sources); \emph{(ii)} the yield of the $\eta$ is fixed from the
Dalitz channel saturating the very low mass part of the spectrum;
\emph{(iii)} the yield of the $\rho$ is fixed by assuming a
$\rho/\omega$ ratio of 1.2 (as expected from previous pp and pBe
measurements and from statistical models). The data show a clear
excess when compared to this cocktail, becoming more pronounced with
increasing centrality.  Thanks to the good mass resolution, the excess
could be isolated by subtraction of the known
sources~\cite{Damjanovic:2006jq} (Fig.~\ref{fig:mass_spectra}, right),
namely the conservative cocktail described above, with the exception
of the $\rho$: since the modification in the spectrum is generally
understood as due to in-medium modification of this particle, it does
not make sense to subtract from the spectrum its vacuum line-shape.
The difference spectrum obtained in this way is a lower limit to the
excess, as a consequence of the procedure adopted to fix yields (for
example, there is no excess at very low masses by construction).
Fig.~\ref{fig:excess_centrality} shows the excess as a function of
centrality. For comparison, the cocktail $\rho$ and $D\bar{D}$ are
also plotted. A clear excess rising with centrality and centered at
the nominal $\rho$ pole is visible. The excess spectrum is robust
against systematical uncertainties: both the combinatorial background
and the fake matches are flat as a function of mass. A 1\% error on
combinatorial background or a 5\% error on fake matches would not
change the overall features of the spectrum. Changing the procedure to
fix the $\eta$ normalization, for example by requiring that it
accounts for 90\% of the data at very low masses, has a small effect
on the lowest part of the spectrum only, while the yield above 0.4~GeV
is unaffected (Fig.~\ref{fig:theory}).

The excess was compared to theoretical models available before the
NA60 data\cite{Rapp:1999ej,Brown:2001nh}. Dropping mass scenarios seem
to be ruled out, while broadening models are strongly favored as can
be seen in the left part of Fig.~\ref{fig:theory} (theoretical models
are normalized to data in the range $M < 0.9~\mathrm{GeV}$). None of
the existing models was able to describe the data in full detail. The
qualitative features of the spectrum are consistent with $\rho$
production via $\pi\pi$ annihilation in the framework of a hadron
many-body theory. The prediction, however, fails to describe the
excess spectrum for $M > 0.9~\textrm{GeV}$. An updated version of this
model\cite{vanHees:2006iv}, implementing an improved fireball dynamics
and $4\pi$ processes (including a contribution from vector-axialvector
mixing) is able to describe the spectrum, even in absolute term
(Fig.~\ref{fig:theory} right, solid curve).  A different model, based
on a similar idea but with a different technical implementation,
considering only a hot pion gas without baryons (which are an
essential ingredient in the previous approach)~\cite{Renk:2006ax} is
also able to describe the data in absolute terms
(Fig.~\ref{fig:theory} right, dashed-dotted curve).  The $4\pi$
processes are not included in this model. Both models require a
continuum contribution from a partonic spectral function, which is
much stronger in the second one and fills the high mass region. A
conclusive discrimination between different scenarios may come from
the study of the \pt~dependence of the excess data.  Attempts to
theoretically understand NA60 data are progressing steadily and more
approaches, different from the ones briefly outlined here, are
appearing in the literature~\cite{moretheory}.

\end{document}